\documentclass{llncs}
\newfont{\mycrnotice}{ptmr8t at 7pt}
\newfont{\myconfname}{ptmri8t at 7pt}
\usepackage{mathptmx}
\usepackage{amssymb}
\usepackage{caption}
\usepackage{graphicx}
\usepackage{epstopdf}
\usepackage{epsfig,graphicx}
\usepackage{color}
\usepackage{url}
\usepackage{tabularx}
\usepackage{xfrac}
\usepackage{array}
\usepackage{subfigure}
\usepackage{multirow}
\usepackage{enumitem}
\usepackage{amsmath}
\usepackage{algorithm,algorithmic}
\usepackage{hhline}
 \usepackage{tikz}
\usepackage{booktabs}
\usepackage[title]{appendix}
\usepackage{adjustbox}
\let\subparagraph\relax

\usepackage{titlesec}
\titlespacing\section{1pt}{12pt plus 2pt minus 1pt}{0pt plus 1pt minus 1pt}
\titlespacing\subsection{1pt}{12pt plus 2pt minus 1pt}{0pt plus 1pt minus 1pt}
\titlespacing\subsubsection{1pt}{12pt plus 2pt minus 1pt}{0pt plus 1pt minus 1pt}

\usepackage{hyperref}
\hypersetup{
     colorlinks   = true,
     citecolor    = gray,
     linkcolor = blue
}
\begin{document}


%


\title{On Early-stage Debunking Rumors on Twitter: Leveraging the Wisdom of Weak Learners}

\author{Tu Nguyen\inst{1} \and Cheng Li\inst{2} \and Claudia Nieder\'ee\inst{1} \\
  \institute{L3S Research Center / Leibniz Universit\"{a}t Hannover \\
  \email{\{tunguyen,niederee\}@L3S.de}
  \and 
  SAP S/4 Hana Cloud Foundation\\
\email{cheng.li@SAP.com}
}}



\maketitle

%
%
\begin{abstract}
Recently a lot of progress has been made in rumor modeling and rumor detection 
for micro-blogging streams. However, existing automated methods do not 
perform very well for early rumor detection, which is crucial in many 
settings, e.g., in crisis situations. One reason for this is that 
aggregated rumor features such as propagation features, which work well 
on the long run, are - due to their accumulating characteristic - not very helpful 
in the early phase of a rumor. In this work, we present an approach for early rumor detection, which leverages Convolutional Neural Networks for learning the hidden representations of individual rumor-related tweets to gain insights on the credibility of each tweets. We then aggregate the predictions from the very beginning of a rumor to obtain the overall event credits (so-called \textit{wisdom}), and finally combine it with a time series based rumor classification model. Our extensive experiments show a clearly improved classification performance within the critical very first hours of a rumor. For a better understanding, we also conduct an extensive feature evaluation that emphasized on the early stage and shows that the low-level credibility has best predictability at all phases of the rumor lifetime. 

\end{abstract}


\vspace{-0.2cm} 
\vspace{-0.2cm}
\vspace{-0.2cm}
\section{Introduction}
\label{sec:intro}
\begin{figure}[t]
\centering
\includegraphics[width=0.6\columnwidth]{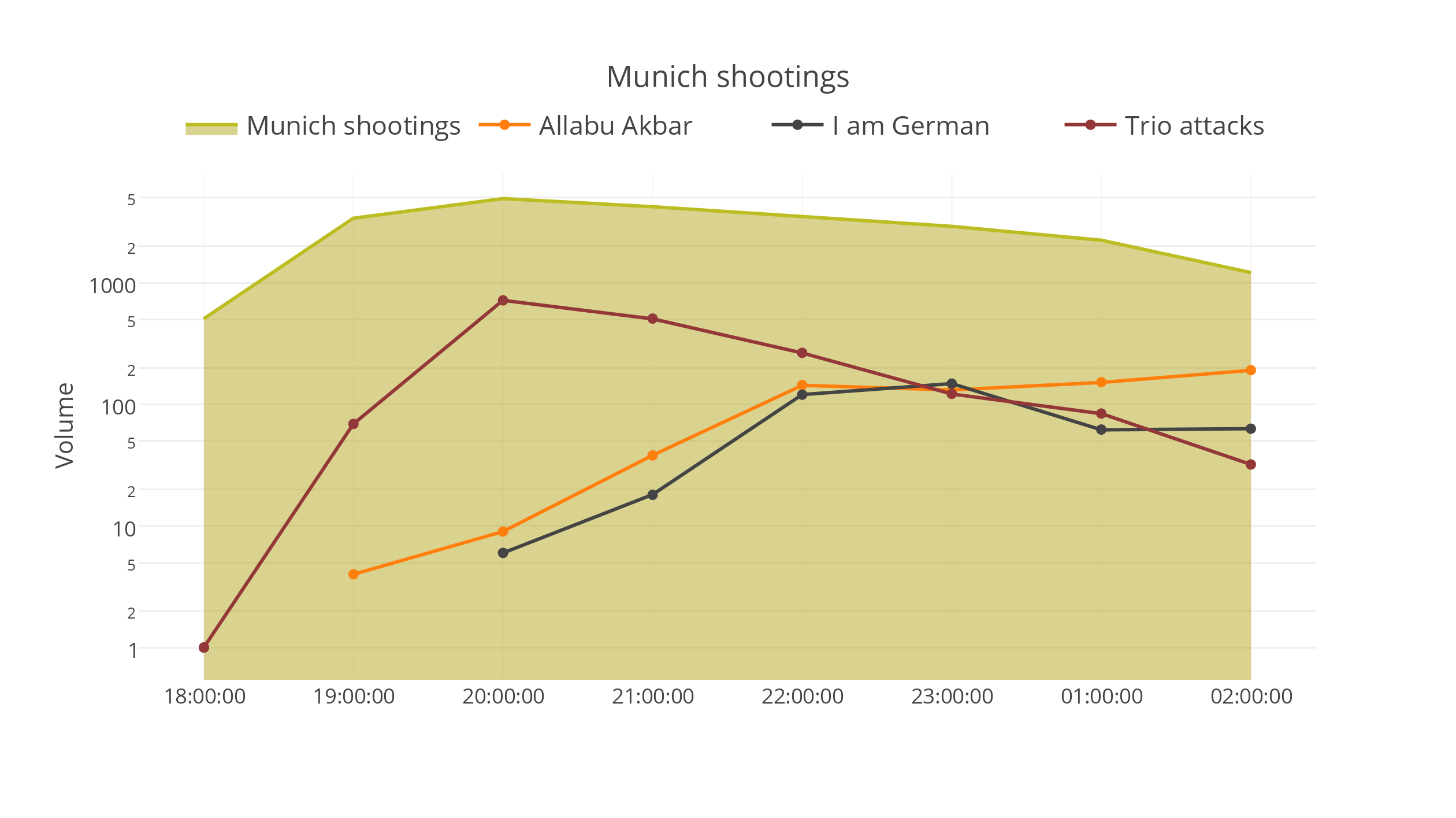}
\vspace{-0.2cm}
\caption{The \textit{Munich shooting} and its sub-events burst after the first 8 hours, y-axis is English tweet volume.}
\label{fig:munich}
\end{figure}

Widely spreading rumors can be harmful to the government, markets and society and reduce the usefulness of social media channel such as Twitter by affecting the reliability of their content. 
Therefore, effective method for detecting rumors on Twitter are crucial and rumors should be detected as early as possible before they widely spread. As an example, let us recall of the shooting incident that happened in the vicinity of the Olympia shopping mall, Munich; in a summer day, 2016.
Due to the unclear situation at early time, numerous rumors about the event 
did appear and they started to circulate very fast over social media.
The city police had to warn the population to refrain from spreading related news on Twitter as it was getting out of control: \textit{``Rumors are wildfires that are difficult to put out and traditional news sources or official channels, such as police departments, subsequently struggle to communicate verified information to the public, as it gets lost under the flurry of false information.''}~\footnote{\scriptsize{Deutsche Welle: \url{http://bit.ly/2qZuxCN}}}
Figure~\ref{fig:munich} shows the rumor \textit{sub-events} in the early stage of the event \textsf{Munich shooting}. The first \textit{terror-indicating} ``news'' --The gunman shouted `Allahu Akbar'-- was widely disseminated on Twitter right after the incident by an unverified account. Later the claim of three gunmen also spread quickly and caused public tension. In the end, all three information items were falsified.  

We follow the rumor definition~\cite{qazvinian2011rumor} considering a rumor (or fake news) as a statement whose truth value is unverified or deliberately false. 
A wide variety of features has been used in existing work in rumor detection 
such as~\cite{castillo2011information,gupta2014tweetcred,jin2013epidemiological,liu2015real,madetecting,ma2015detect,mendoza2010twitter,wu2015false,yang2012automatic}. Network-oriented and other aggregating features such 
as propagation pattern have proven to be effective for this task. 
Unfortunately, the inherently accumulating characteristic of such features, 
which require some time (and Twitter traffic) to mature, does not 
make them very apt for early rumor detection. A first semi-automatic approach focussing 
on early rumor detection presented by Zhao et al.~\cite{zhao2015enquiring}, 
thus, exploits rumor signals such as enquiries that might already arise 
at an early stage. Our fully automatic, cascading rumor detection method follows 
the idea on focusing on early rumor signals on text contents; which is the most reliable source before the rumors widely spread. Specifically, we learn a more complex representation of single tweets using Convolutional Neural Networks, that could capture more hidden meaningful signal than only enquiries to debunk rumors. 
~\cite{tong2017,madetecting} also use RNN for rumor debunking. However, in their work, RNN is used at \textit{event-level}. The classification leverages only the deep data representations of aggregated tweet contents of the whole event, while ignoring exploiting other --in latter stage--effective features such as user-based features and propagation features. Although, tweet contents are merely the only reliable source of clue at early stage, they are also likely to have doubtful perspectives and different stands in this specific moment. In addition, they could relate to rumorous sub-events (see e.g., the \textsf{Munich shooting}). Aggregating all relevant tweets of the event at this point can be of noisy and harm the classification performance. One could think of a sub-event detection mechanism as a solution, however, detecting sub-events at real-time over Twitter stream is a challenging task~\cite{meladianos2015degeneracy}, which  increases latency and complexity. In this work, we address this issue by deep neural modeling only at single tweet level. Our intuition is to leverage the ``wisdom of the crowd'' theory; such that even a certain portion of tweets at a moment (mostly early stage) are weakly predicted (because of these noisy factors), the ensemble of them would attribute to a stronger prediction. 

In this paper, we make the following contributions with respect to rumor detection:
\vspace{-0.2cm} 
\begin{itemize}
	\item We develop a machine learning approach for modeling tweet-level credibility. Our CNN-based model reaches 81\% accuracy for this novel task, that is even hard for human judgment. The results are used to debunk rumors in an ensemble fashion.

 	\item Based on the credibility model we develop a novel and effective cascaded model for rumor classification. The model uses time-series structure of features to capture their temporal dynamics. Our model clearly 
outperforms strong baselines, especially for the targeted early stage of the diffusion. It already
reaches over 80\% accuracy in the first hour going up to over 90\% accuracy over time.

 \end{itemize}
 \vspace{-0.2cm}

\vspace{-0.2cm}
\section{Related Work}
\label{sec:rw}

A variety of issues have been investigated using data, structural information, and the dynamics of the microblogging 
platform Twitter including event detection \cite{kimmey2015twitter}, spam detection ~\cite{ahmed2012mcl,wang2010don}, or sentiment detection~\cite{barbosa2010robust}. Work on rumor detection in Twitter is 
less deeply researched so far, although rumors and their 
spreading have already been investigated for a long time in psychology ~\cite{allport1947psychology,borge2012absence,sunstein2014rumors}.
Castillo et al. researched the information credibility on Twitter\cite{castillo2011information,gupta2014tweetcred}. The work, however, is based solely on people's attitude (trustful or not) to a tweet not the credibility of the tweet itself. In other words, a false rumor tweet can be trusted by a reader, but it might anyway contain false information. The work still provides a good start of researching rumor detection. 

Due to the importance of information propagation for rumors and their detection, there are also different simulation studies~\cite{seo2012identifying,tripathy2010study} about rumor propagations on Twitter. Those works provide relevant insights, but such simulations cannot fully reflect the complexity of real networks. Furthermore, there are recent work on propagation modeling based on epidemiological methods~\cite{bao2013new,jin2013epidemiological,kwon2013prominent}, yet over a long studied time, hence how the propagation patterns perform at early stage is unclear. Recently, ~\cite{wu2015false} use unique features of Sina Weibo to study the propagation patterns and achieve good results. Unfortunately Twitter does not give such details of the propagation process as Weibo, so these work cannot be fully applied to Twitter. 

Most relevant for our work is the work presented in~\cite{ma2015detect}, where a time series model to capture the time-based variation of social-content features is used. We build upon the idea of their \textit{Series-Time Structure}, when building our approach for early rumor detection with our extended dataset, and we provide a deep analysis on the wide range of features change during diffusion time. Ma et al.~\cite{madetecting} used Recurrent Neural Networks for rumor detection, they batch tweets into time intervals and model the time series as a RNN sequence. Without any other handcrafted features, they got almost 90\% accuracy for events reported in Snope.com. As the same disadvantage of all other deep learning models, the process of learning is a black box, so we cannot envisage the cause of the good performance based only on content features. The model performance is also dependent on the tweet retrieval mechanism, of which quality is uncertain for stream-based trending sub-events.



\section{Single Tweet Credibility Model} 

\begin{figure}[t]
\centering
\includegraphics[width=0.7\columnwidth]{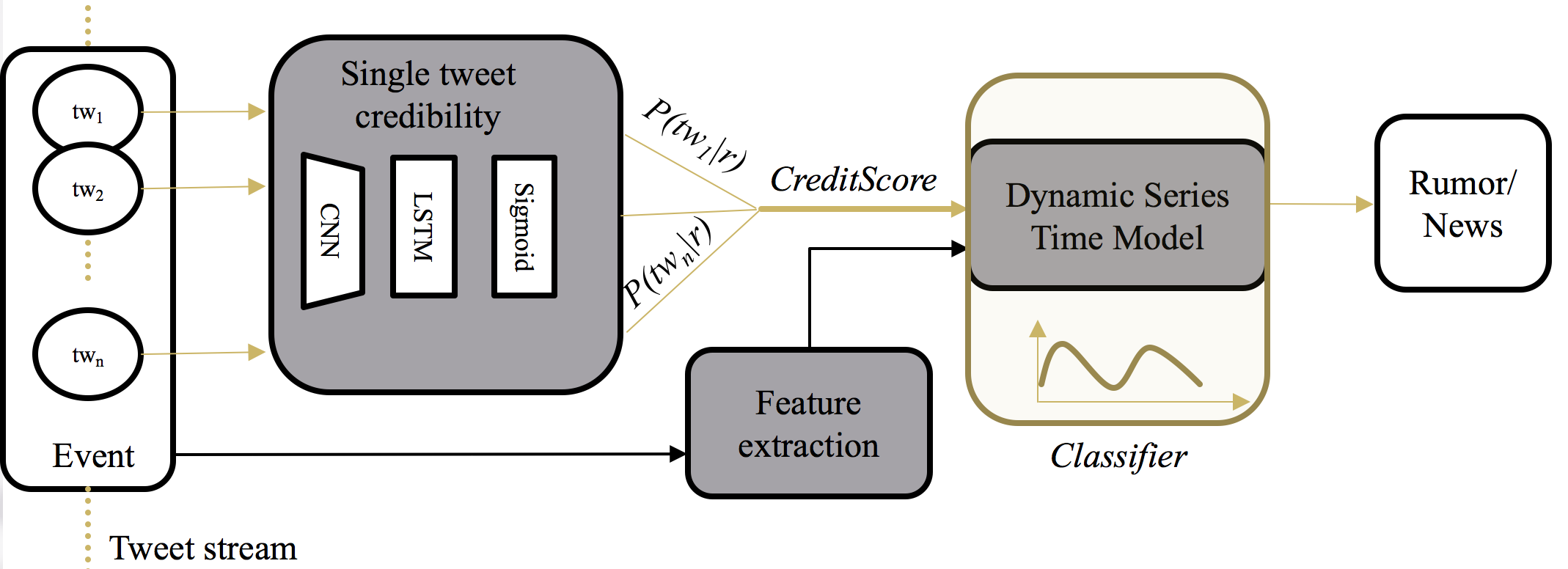}
\caption{Pipeline of our rumor detection approach.}
\label{fig:pipeline}
\end{figure}

Before presenting our Single Tweet Credibility Model, we will start 
with an overview of our overall rumor detection method. 
The processing pipeline of our classification approach is shown in Figure \ref{fig:pipeline}. In the first step, relevant tweets for an event are gathered. Subsequently, in the upper part of the pipeline, 
we predict tweet credibilty with our pre-trained credibility model and aggregate the prediction probabilities on single tweets (CreditScore).
In the lower part of the pipeline, we extract features from tweets and combine them with the creditscore to construct the feature vector in a time series structure called Dynamic Series Time Model. These feature vectors are used to train the classifier for rumor vs. (non-rumor) news  classification.

Early in an event, the related tweet volume is scanty and there are no clear propagation pattern yet. For the credibility model we, therefore, leverage the signals derived from tweet contents. Related work often uses aggregated content~\cite{liu2015real,ma2015detect,zhao2015enquiring}, since individual tweets are often too short and contain slender context to draw a conclusion. However, content aggregation is problematic for hierarchical events and especially at early stage, in which tweets are likely to convey doubtful and contradictory perspectives. Thus, a mechanism for carefully considering the `vote' for individual tweets is required. In this work, we overcome the restrictions (e.g., semantic sparsity) of traditional text representation methods (e.g., bag of words) in handling short text by learning low-dimensional tweet embeddings. In this way, we achieve a rich hidden semantic representation for a more effective classification.
\vspace{-0.2cm}
\subsection{Exploiting Convolutional and Recurrent Neural Networks}
\label{subsec:rnn}

\begin{figure}[t]
\centering
\includegraphics[width=0.7\columnwidth]{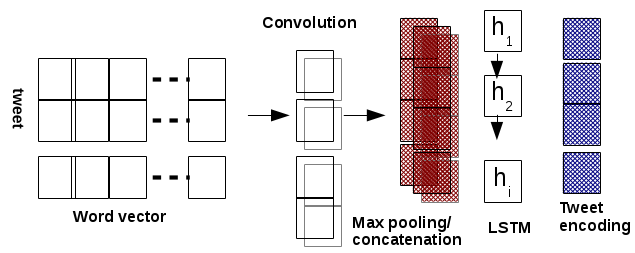}
\caption{CNN+LSTM for tweet representation.}
\label{fig:cnnlstm}
\end{figure}

Given a tweet, our task is to classify whether it is associated with either a news or rumor. Most of the previous work~\cite{castillo2011information,gupta2014tweetcred} on tweet level only aims to measure the \textit{trustfulness} based on human judgment (note that even if a tweet is trusted, it could anyway relate to a rumor). Our task is, to a point, a reverse engineering task; to measure the probability a tweet refers to a \textit{news} or \textit{rumor} event; which is even trickier. We hence, consider this a weak learning process. Inspired by \cite{zhou2015c}, we combine CNN and RNN into a unified model for tweet representation and classification. The model utilizes CNN to extract a sequence of higher-level phrase representations, which are fed into a long short-term memory (LSTM) RNN to obtain the tweet representation. This model, called CNN+RNN henceforth, is able to capture both local features of phrases (by CNN) as well as global and temporal tweet semantics (by LSTM)(see Figure~\ref{fig:cnnlstm}).

\textbf{Representing Tweets:} 
Generic-purpose tweet embedding in~\cite{dhingra2016tweet2vec,vosoughi2016tweet2vec} use character-level RNN to represent tweets that in general, are noisy and of idiosyncratic nature. We discern that tweets for rumors detection are often triggered from professional sources. Hence, they are linguistically clean, making word-level embedding become useful. In this work, we do not use the pre-trained embedding (i.e., \textit{word2vec}), but instead learn the word vectors from scratch from our (large) rumor/news-based tweet collection. The effectiveness of fine-tuning by learning task-specific word vectors is backed by~\cite{kim2014convolutional}. We represent tweets as follows: Let $x_{i} \in \mathcal{R}$ be the $k$-dimensional word vector corresponding to the $i$-th word in the tweet. A tweet of length $n$ (padded where necessary) is represented as: $
x_{1:n} = x_{1} \oplus x_{2} \oplus \cdots \oplus x_{n}
$, where $\oplus$ is the concatenation operator. In general, let $x_{i:i+j}$ refer to the concatenation of words $x_{i},x_{i+1},...,x_{i+j}$. A convolution operation involves a filter $w \in \mathcal{R}^{hk}$, which is applied to a window of $h$ words to produce a feature. For example, a feature $c_{i}$ is generated from a window of words $x_{i:i+h-1}$ by: $ c_{i} = f(w \cdot x_{i:i+h-1} + b)$.

Here $b \in \mathcal{R}$ is a bias term and $f$ is a non-linear function such as the hyperbolic tangent. This filter is applied to each possible window of words in the tweet $\{x_{1:h}, x_{2:h+1},...,x_{n-h+1:n}\}$ to produce a feature map: $c = [c_{1},c_{2},...,c_{n-h+1}]$ with $c \in \mathcal{R}^{n-h+1}$. A max-over-time pooling or dynamic k-max pooling is often applied to feature maps after the convolution to select the most or the k-most important features. We also apply the  1D max pooling operation over the time-step dimension to obtain a fixed-length output.


\textbf{Using Long Short-Term Memory RNNs:} RNN are able to propagate historical information via a chain-like neural network architecture. While processing sequential data, it looks at the current input $x_{t}$ as well as the previous output of hidden state $h_{t-1}$ at each time step. The simple RNN hence has the ability to capture context information. However, the length of reachable context is often limited. The gradient tends to vanish or blow up during the back propagation. To address this issue, LSTM was introduced in~\cite{hochreiter1997long}. The LSTM architecture has a range of repeated modules for each time step as in a standard RNN. At each time step, the output of the module is controlled by a set of gates in $\mathcal{R}^{d}$ as a function of the old hidden state $h_{t-1}$ and the input at the current time step $x_{t}$: forget gate $f_{t}$, input gate $i_{t}$, and output gate $o_{t}$. 

\subsection{CNN+LSTM for tweet-level classification.}
We regard the output of the hidden state at the last step of LSTM as the final tweet representation and we add a softmax layer on top. We train the entire model by minimizing the cross-entropy error. Given a training tweet sample $x^{(i)}$, its true label $y_{j}^{(i)} \in \{y_{rumor},y_{news}\}$ and the estimated probabilities $\tilde{y}_{j}^{(i)} \in [0..1]$ for each label $j \in \{rumor,news\}$, the error is defined as:

\begin{equation}
\mathsf{L}(x^{(i)},y^{(i)}) = 1\{y^{(i)}=y_{rumor}\}log(\tilde{y}_{rumor}^{(i)}) + 1\{y^{(i)}=y_{news}\}log(\tilde{y}_{news}^{(i)})
\end{equation}

where $1$ is a function converts boolean values to $\{0,1\}$. We employ stochastic gradient descent (SGD) to learn the model parameters.

\section{Time Series Rumor Detection Model} 
\label{sec:timr_seriers_rumor_model}
As observed in~\cite{madetecting,ma2015detect}, rumor features are very prone to change during an event's development. In order to capture these temporal variabilities, we build upon the Dynamic Series-Time Structure (DSTS) model (time series for short) for feature vector representation proposed in~\cite{ma2015detect}. We base our credibility feature on the time series approach and train the classifier with features from diffent high-level contexts (i.e., users, Twitter and propagation) in a cascaded manner. In this section, we first detail the employed Dynamic Series-Time Structure, then describe the high and low-level ensemble features used for learning in this pipeline step.

  \subsection{ Dynamic Series-Time Structure (DSTS) Model} 

%

For an event $E_i$ we define a time frame given by $timeFirst_i$ as the start time of the event and $timeLast_i$ as the time of the last tweet of the event in the observation time.
We split this event time frame into N intervals and associate each tweet to one of the intervals according to its creation time.  Thus, we can generate a vector V($E_i$) of features for each time interval. In order to capture the changes of feature over time, we model their differences between two time intervals. So the model of DSTS is represented as: $V(E_i)=(\textbf{F}^D_{i,0}, \textbf{F}^D_{i,1},..., \textbf{F}^D_{i,N},\textbf{S}^D_{i,1},..., \textbf{S}^D_{i,N})$, where $\textbf{F}^D_{i,t}$ is the feature vector in time interval t of event $E_i$. $\textbf{S}^D_{i,t}$ is the difference between $\textbf{F}^D_{i,t}$ and $\textbf{F}^D_{i,t+1}$. V($E_i$ ) is the time series feature vector of the event $E_i$. $\textbf{F}^D_{i,t}=(\widetilde{ f}_{i,t,1},\widetilde{ f}_{i,t,2},...,\widetilde{ f}_{i,t,D})$. And $\textbf{S}^D_{i,t}=\frac{\textbf{F}^D_{i,t+1}-\textbf{F}^D_{i,t}}{Interval(E_i)}$. We use Z-score to normalize feature values;
$
\widetilde{f}_{i,t,k}=\frac{f_{i,t+1,k}-\overline{f}_{i,k}}{\sigma(f_{i,k})}
$
 where $f_{i,t,k}$ is the k-th feature of the event $E_i$ in time interval t. The mean of the feature k of the event $E_i$ is denoted as  $\overline{f}_{i,k}$ and $\sigma(f_{i,k})$ is the standard deviation of the feature k over all time intervals. We can skip this step, when we use Random Forest or Decision Trees, because they do not require feature normalization.


%

\subsection{Features for the Rumor Detection Model} 
\label{sub:features}

In selecting features for the rumor detection model, we have followed 
two rationales: a) we have selected features that we expect to be useful 
in early rumor detection and b) we have collected a broad 
range of features from related work as a basis for investigating 
the time-dependent impact of a wide variety of features 
in our time-dependence study. 
In total, we have constructed over 50 features\footnote{details are listed in the Appendix.}
in the three main categories i.e., \textit{Ensemble}, \textit{Twitter} and \textit{Epidemiological} features.  
We refrained from using network features, since they are  expected to 
be of little use in early rumor detection~\cite{DBLP:journals/corr/ContiLLLQ17}, since user networks around events 
need time to form. Following our general idea, none of our features are extracted from the content aggregations. Due to space limitation, we describe only our main features as follows.
 
 \subsubsection{Ensemble Features.} 
 \label{sub:ef} \indent
 We consider two types of Ensemble Features: features accumulating crowd wisdom and averaging feature for the Tweet credit Scores. 
The former are extracted from the surface level while the latter comes from the low dimensional level of tweet embeddings; that in a way augments the sparse crowd at early stage.

\textbf{\textit{CrowdWisdom:}} Similar to~\cite{liu2015real}, the core idea is to leverage the public's common sense for rumor detection: If there are more people denying or doubting the truth of an event, this event is more likely to be a rumor. For this purpose, ~\cite{liu2015real} use an extensive list of bipolar sentiments with a set of combinational rules. In contrast to mere \textit{sentiment} features, this approach is more tailored \textit{rumor} context (difference not evaluated in~\cite{liu2015real}). We simplified and generalized the ``dictionary'' by keeping only a set of carefully curated \textit{negative words}. We call them ``debunking words" e.g., \textit{hoax}, \textit{rumor} or \textit{not true}. Our intuition is, that the attitude of doubting or denying events is in essence sufficient to distinguish rumors from news. What is more, this generalization augments the size of the crowd (covers more 'voting' tweets), which is crucial, and thus contributes to the quality of the crowd wisdom. In our experiments, ``debunking words" is an high-impact feature, but it needs substantial time to ``warm up"; that is explainable as the crowd is typically sparse at early stage. 

\textbf{\textit{CreditScore:}}
The sets of single-tweet models' predicted probabilities are combined using an \textit{ensemble averaging}-like technique. In specific, our pre-trained $CNN+LSTM$ model predicts the credibility of each tweet $tw_{ij}$ of event $E_{i}$. The \textit{softmax} activation function outputs probabilities from 0 (rumor-related) to 1 (news). Based on this, we calculate the average prediction probabilities of all tweets $tw_{ij} \in E_{i}$ in a time interval $t_{ij}$. In theory there are different sophisticated ensembling approaches for averaging on both training and test samples; but in a real-time system, it is often convenient (while effectiveness is only affected marginally) to cut corners. In this work, we use a sole training model to average over the predictions. We call the outcome \textsf{CreditScore}. 

\vspace{-0.2cm}




%
%
%
%
%

\vspace{-0.2cm}
\section{Experimental Evaluation} 
  \subsection{Data Collection} 
  \label{sec:dataset_single}
To construct the training dataset, we collected rumor stories from online rumor tracking websites such as \textbf{snopes.com} and \textbf{urbanlegends.about.com}. In more detail, we crawled 4300 stories from these websites. From the 
story descriptions we manually constructed queries to retrieve the relevant tweets for 270 rumors with high impact. Our approach to query construction mainly follows~\cite{gupta2014tweetcred}. For the news event instances (non-rumor examples), we make use of the manually constructed corpus from Mcminn et al.~\cite{mcminn2013building}, which covers 500 real-world events. In~\cite{mcminn2013building}, tweets are retrieved via Twitter firehose API from $10^{th}$ of October 2012 to $7^{th}$ of November 2012. The involved events are manually verified and relate to tweets with relevance judgments, which results in a high quality corpus. From the 500 events, we select top 230 events with the highest tweet volumes (as a criteria for event impact). Furthermore, we have added 40 other news events, which happened around the time periods of our rumors. This results in a dataset of 270 rumors and 270 events.  The dataset details are shown in Table \ref{tab:Tweet_Volume}. To serve our learning task. we then constructs two distinct datasets for (1) single tweet credibility and (2) rumor classification.

\begin{table}[t]
 \centering
\scalebox{0.8}{
 \begin{tabular}{@{}cccccc@{}}
 \toprule
 \textbf{Type} & \textbf{Min Volume} & \textbf{Max Volume}& \textbf{Total} &\textbf{Average} \\ \midrule
 News & 98 & 17414 & 345235 & 1327.82 \\ \midrule
 Rumors & 44  & 26010& 182563 & 702.06\\
 	 \bottomrule
 \end{tabular}}
 \caption{Tweet Volume of News and Rumors}
 \label{tab:Tweet_Volume}
\end{table}
 \textbf{Training data for single tweet classification.} Here we follow our assumption that an event might include sub-events for which relevant tweets are rumorous. To deal with this complexity, we train our single-tweet learning model only with manually selected \textit{breaking and subless~\footnote{the terminology \textit{subless} indicates an event with no sub-events for short.}} events from the above dataset. In the end, we used 90 rumors and 90 news associated with 72452 tweets, in total. This results in a highly-reliable large-scale ground-truth of tweets labelled as \textit{news}-related and \textit{rumor}-related, respectively. Note that the labeling of a tweet is inherited from the event label, thus can be considered as an semi-automatic process. 

\subsection{Single Tweet Classification Experiments} 
For the evaluation, we developed two kinds of classification models: traditional classifier with handcrafted features and neural networks without tweet embeddings. For the former, we used 27 distinct surface-level features extracted from single tweets (analogously to the Twitter-based features presented in Section~\ref{sub:features}). For the latter, we select the baselines from NN-based variations, inspired by state-of-the-art short-text classification models, i.e., Basic tanh-RNN~, 1-layer GRU-RNN, 1-layer LSTM, 2-layer GRU-RNN, FastText~\cite{joulin2016bag} and CNN+LSTM~\cite{zhou2015c} model. The hybrid model CNN+LSTM is adapted in our work for tweet classification. 

%
%
\textbf{Single Tweet Model Settings.} For the evaluation, we shuffle the 180 selected events and split them into 10 subsets which are used for 10-fold cross-validation (we make sure to include near-balanced folds in our shuffle). We implement the 3 non-neural network models with Scikit-learn\footnote{scikit-learn.org/}. Furthermore, neural networks-based models are implemented with TensorFlow~\footnote{https://www.tensorflow.org/} and Keras\footnote{https://keras.io/}. The first hidden layer is an embedding layer, which is set up for all tested models with the embedding size of 50. The output of the embedding layer are low-dimensional vectors representing the words. To avoid overfitting, we use the 10-fold cross validation and dropout for regularization with dropout rate of 0.25.

\textbf{Single Tweet Classification Results.}  The experimental results of are shown in Table \ref{tab:single_result}. The best performance is achieved by the CNN+LSTM model with a good accuracy of 81.19\%. The non-neural network model with the highest accuracy is RF. However, it reaches only 64.87\% accuracy and the other two non-neural models are even worse. So the classifiers with hand-crafted features are less adequate to accurately distinguish between rumors and news. 
 
 
\begin{minipage}{\columnwidth}
  \begin{minipage}[b]{0.45\columnwidth}
      \centering
    \begin{adjustbox}{center, width=0.8\columnwidth} 
   \begin{tabular}{@{}lllllll@{}}
 \toprule
 \textbf{Model} & \textbf{Accuracy} \\ \midrule
  \textbf{CNN+LSTM} & \textbf{0.8119 }\\
 2-layer GRU & 0.7891\\
 1-layer GRU & 0.7644\\
 1-layer LSTM & 0.7493\\
 Basic RNN with tanh &  0.7291\\
 FastText &  0.6602\\ \bottomrule
 Random Forest & \textbf{0.6487 }\\
 SVM &  0.5802\\
 Decision Trees &  0.5774\\ \bottomrule
 \end{tabular}
 \end{adjustbox}
\captionof{table}{Single Tweet Classification Performance}
\vspace{1cm}
 \label{tab:single_result}

 \end{minipage}
  \hfill
  \begin{minipage}[b]{0.45\columnwidth}
    \centering
    \begin{adjustbox}{center, width=0.7\columnwidth}   
 \begin{tabular}{@{}lllllll@{}}
\toprule
\textbf{Feature} & \textbf{Importance} \\ \midrule
PolarityScores	&	0.146\\
Capital	&	0.096\\
LengthOfTweet  &	0.092\\
UserTweets  &	0.087 \\
UserFriends  &	0.080 \\
UserReputationScore  &	0.080 \\
UserFollowers   &	0.079 \\
NumOfChar	&	0.076\\
Stock	&	0.049\\
NumNegativeWords	&	0.030\\
Exclamation	&	0.023\\
\bottomrule

\end{tabular}
\end{adjustbox}
 \captionof{table}{Top Features Importance}
 \vspace{1cm}
\label{tab:Features_Importance}

\end{minipage}
 \end{minipage}

\textbf{Discussion of Feature Importance}  
For analyzing the employed features, we rank them by importances using RF (see \ref{tab:Features_Importance}). The best feature is related to sentiment polarity scores. There is a big difference between the sentiment associated to rumors and the sentiment associated to real events in relevant tweets. In specific, the average polarity score of news event is -0.066 and  the average of rumors is -0.1393, showing that rumor-related messages tend to contain more negative sentiments. Furthermore, we would expect that verified users are less involved in the rumor spreading. However, the feature appears near-bottom in the ranked list, indicating that it is not as reliable as expected. Also interestingly, ``IsRetweeted'' feature is pretty much useless, which means the probability of people retweeting rumors or true news are similar (both appear near-bottom in the ranked feature list).

It has to be noted here that even though we obtain reasonable results on the classification task in general, the prediction performance varies considerably along the time dimension. This is understandable, since tweets become more distinguishable, only when the user gains more knowledge about the event. 
 \subsection{Rumor Datasets and Model Settings} 
  We use the same dataset described in Section \ref{sec:dataset_single}. In total --after cutting off 180 events for pre-training single tweet model -- our dataset contains 360 events and 180 of them are labeled as rumors. Those rumors and news fall comparatively evenly in 8 different categories, namely \textit{Politics}, \textit{Science}, \textit{Attacks}, \textit{Disaster}, \textit{Art}, \textit{Business}, \textit{Health} and \textit{Other}. Note, that the events in our training data are not necessarily subless, because it is natural for high-impact events (e.g., \textit{Missing MH370} or \textit{Munich shooting}) to contain sub-events. Actually, we empirically found that roughly 20\% of our events (mostly news) contain sub-events.  As a rumor is often of a long circulating story~\cite{friggeri2014rumor}, this results in a rather long time span. In this work, we develop an event identification strategy that focuses on the first 48 hours after the rumor is peaked. We also extract 11,038 domains, which are contained in tweets in this 48 hours time range.

\textbf{Rumor Detection Model Settings.} For the time series classification model, we only report the best performing classifiers, SVM and Random Forest, here. The parameters of SVM with RBF kernel are tuned via grid search to $C=3.0$, $\gamma = 0.2$. For Random Forest, the number of trees is tuned to be 350. All models are trained using 10-fold cross validation. 

 \subsection{Rumor Classification Results} 
We tested all models by using 10-fold cross validation with the same shuffled sequence. The results of these experiments are shown in Table \ref{tab:time_result}. Our proposed model (\textbf{\textit{Ours}}) is the time series model learned with Random Forest including all ensemble features; $TS-SVM$ is the baseline from~\cite{ma2015detect}, and $TS-SVM_{all}$ is the $TS-SVM$ approach improved by using our feature set. In the lower part of the table, $RNN_{el}$ is the RNN model at event-level~\cite{madetecting}. As shown in the Table \ref{tab:time_result} and as targeted by our early detection approach, our model has the best performance in all case over the first 24 hours, remarkably outperforming the baselines in the first 12 hours of spreading. The performance of $RNN_{el}$ is relatively low, as it is based on aggregated \textit{contents}. This is expected as the news (non-rumor) dataset used in ~\cite{madetecting} are crawled also from snopes.com, in which events are often of small granularity (aka. subless). As expected, exploiting contents solely at event-level is problematic for high-impact, evolving events on social media. We leave a deeper investigation on the sub-event issue to future work.

\begin{table}[H]
\centering
\scalebox{0.9}{
\begin{tabular}{|c|ccccccccc|}
\hline
\multicolumn{1}{|c|}{\multirow{2}{*}{Model}} & \multicolumn{9}{c|}{Accuracy in hours}                     \\ \cline{2-10} 
\multicolumn{1}{|l|}{}& 1 & 6 &12& 18 & 24 & 30 & 36  & 42 & \multicolumn{1}{c|}{48} \\\hline
 $Ours$  & \textbf{0.82} &  \textbf{0.84} &  \textbf{0.84} &\underline{0.84} & \textbf{0.87}& \underline{0.87} &\textbf{0.88}&\underline{0.89}&\textbf{0.91 }\\
$TS-SVM_{all}$ &\underline{0.76} & 0.79  &  \underline{0.83} & 0.83 &   \underline{0.87} &   \textbf{0.88}&   0.86&   0.89 & \underline{0.90}  \\

$TS-SVM_{Credit}$& 0.73 & \underline{0.80}  & 0.83  & \textbf{0.85} &   0.85 & 0.86  &\underline{0.88} &\textbf{0.90} &\underline{0.90}\\

$TS-SVM$~\cite{ma2015detect} &0.69  &0.76 & 0.81& 0.81 &  0.84   &0.86 & 0.87& 0.88 & 0.88  \\
\hline
$RNN_{el}$~\cite{madetecting}&0.68 & 0.77 & 0.81  & 0.81 &  0.84 &0.83  &0.81 &0.85 & 0.86\\
$SVM_{static}+Epi$~\cite{jin2013epidemiological} &0.60& 0.69  & 0.71&0.72   &  0.75 &0.78& 0.75&0.78&0.81 \\   

$SVM_{static}+SpikeM$~\cite{kwon2013prominent}&0.58& 0.68  & 0.72&0.73   &  0.77 &0.78& 0.78&0.79&0.77 \\   

$SVM_{static}$~\cite{yang2012automatic} &0.62& 0.70  & 0.70&0.72   &  0.75 & 0.80& 0.79&0.78&0.77 \\   

\bottomrule           
\end{tabular}
}
 \caption{Performance of different models over time (bold for best accuracy, underlined for second-to-best). \textsf{TS} indicates time-series structure; we separate the TS models (upper) with the static ones (lower).}
 \label{tab:time_result}
\end{table} 

\subsubsection{CreditScore and CrowdWisdom}
\begin{figure}[H]
\centering
\includegraphics[width=0.55\columnwidth]{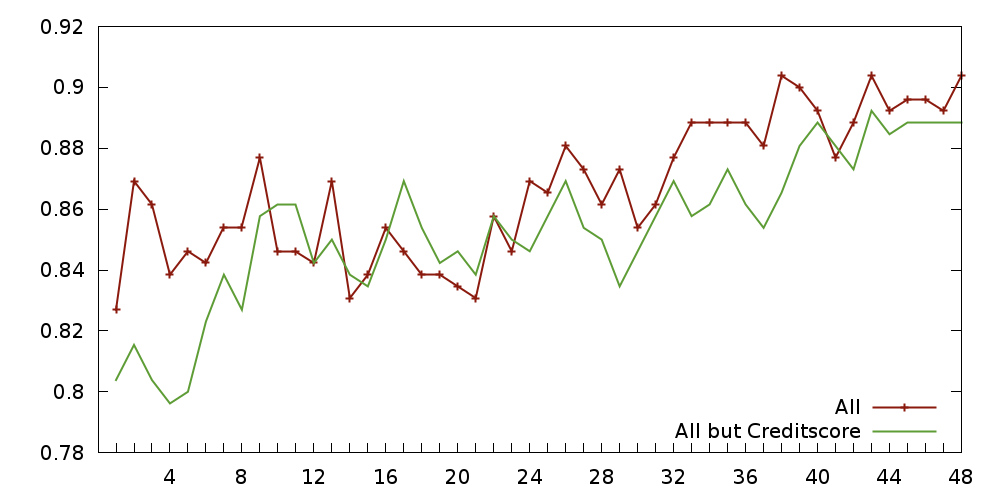}
\caption{Accuracy: All features with and without CreditScore.}
\label{fig:WCVSAF}
\end{figure}

. As shown in Table~\ref{tab:Rank_Credit}, \emph{CreditScore} is the best feature in overall. In Figure \ref{fig:WCVSAF} we show the result of models learned with the full feature set with and without \emph{CreditScore}. Overall, adding \textit{CreditScore} improves the performance, especially for the first 8-10 hours. The performance of \textit{all-but-CreditScore} jiggles a bit after 16-20 hours, but it is not significant. \emph{CrowdWisdom} is also a good feature which can get 75.8\% accuracy as a single feature. But its performance is poor (less than 70\%) in the first 32 hours getting better over time (see Table~\ref{tab:Rank_Credit}). Table~\ref{tab:Rank_Credit} also shows the performance of \textit{sentiment} feature (\textit{PolarityScores}), which is generally low. This demonstrates the effectiveness of our \textit{curated} approach over the \textit{sentiments}, yet the crowd needs time to unify their views toward the event while absorbing different kinds of information. 

%
  
  \begin{table}[H]
\centering
\scalebox{0.8}{
\begin{tabular}{@{\textbf{ }}ccccccccccccccccc@{}}
\toprule
\textbf{Features} & \multicolumn{10}{c}{\textbf{Ranks}} \\\hline
Hours & 1 & 6 & 12 & 18&24&30&36&42&48 & AVG\\\hline
CreditScore & 1 & 0 & 0& 0 & 0&0&0&0&0&0.08\\
CrowdWisdom	 & 34& 38 & 21 & 14& 8& 5& 5& 2&2&13.18\\
PolarityScores & 12 & 15& 23 & 28 & 33& 33& 34& 31&32&28\\
 \bottomrule
 \end{tabular}}
 \caption{Importance ranking of CreditScore, CrowdWisdom and PolarityScores over time; 0 indicates the best rank.}
\label{tab:Rank_Credit}
 \end{table}

\subsubsection{Case Study: Munich Shooting}
\begin{figure*}[h!]
\centering
\subfigure[CreditScore first 12 hours]{\label{fig:munichattackCS}
\centering
\includegraphics[width=0.35\columnwidth]{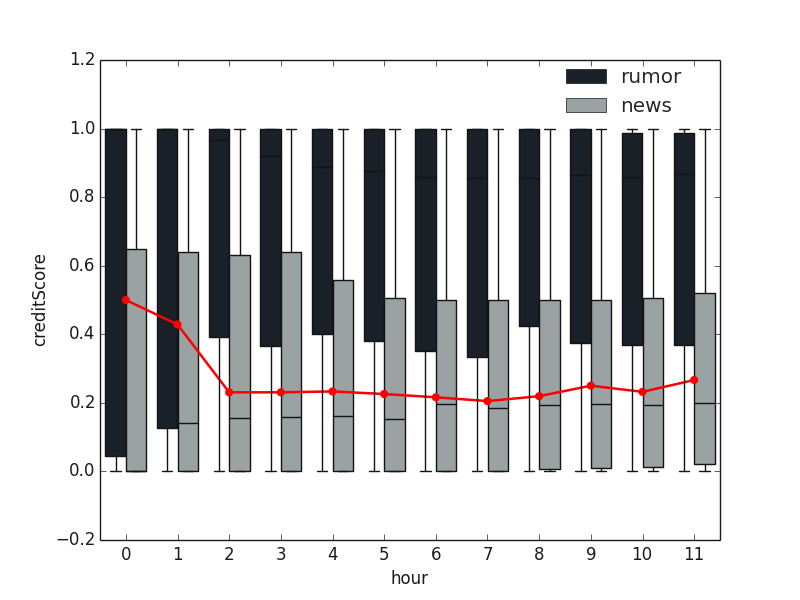}
}
\subfigure[ContainsNews first 12 hours]{\label{fig:munichattackNews}
\centering
\includegraphics[width=0.35\columnwidth]{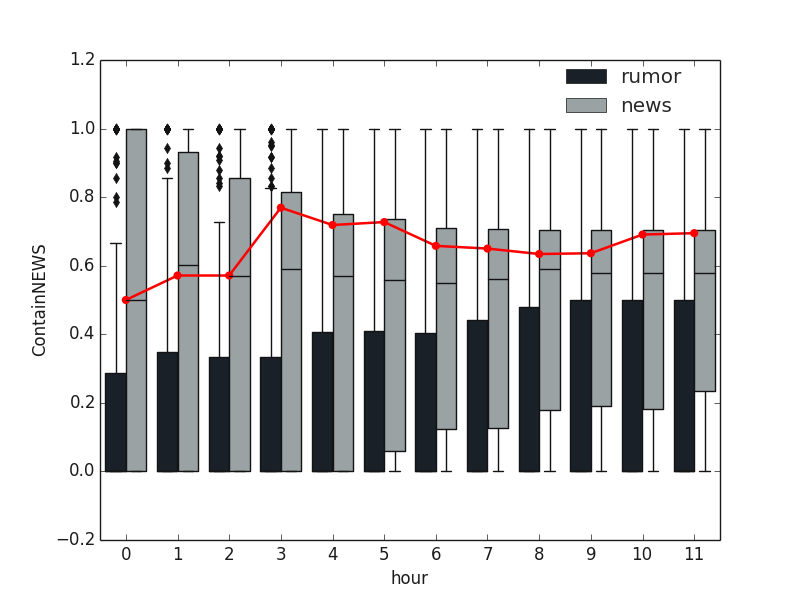}
}
\vskip\baselineskip
\subfigure[CreditScore 48 hours]{\label{fig:munichattackCS2}
\centering
\includegraphics[width=0.35\columnwidth]{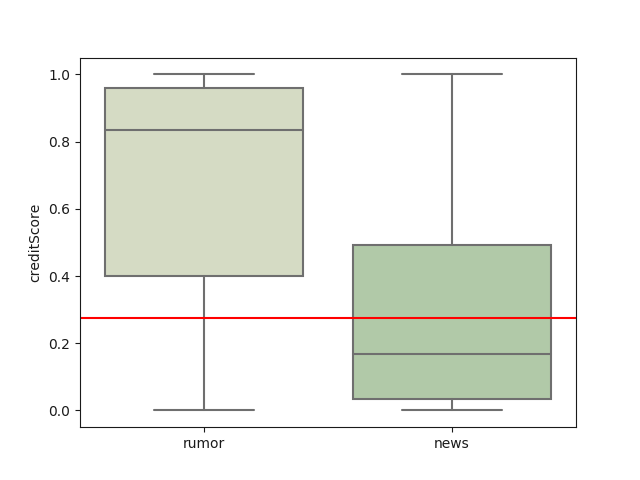}
}
\subfigure[ContainsNews 48 hours]{\label{fig:munichattackNews2}
\centering
\includegraphics[width=0.35\columnwidth]{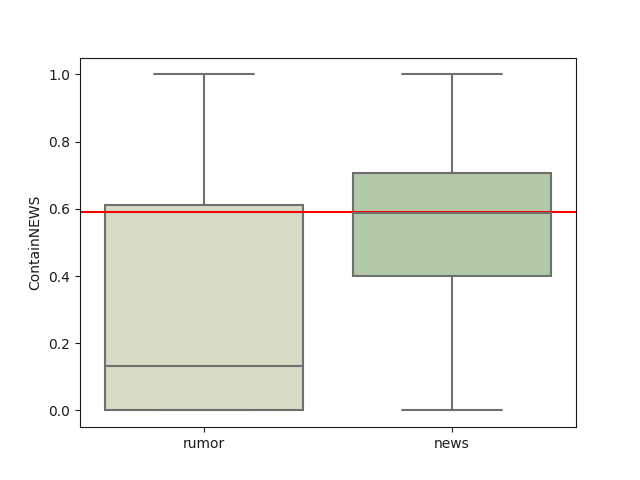}
}
\caption{Creditscore and ContainsNews for \textit{Munich shooting} in red lines, compared with the corresponding average scores for \textit{rumor} and \textit{news}.}
\end{figure*}

. We showcase here a study of the Munich shooting. We first show the event timeline at an early stage. Next we discuss some examples of  misclassifications by our ``weak'' classifier and show some analysis on the strength of some highlighted features. The rough event timeline looks as follows.
\begin{small}
\begin{itemize}
\item At 17:52 CEST, a shooter opened fire in the vicinity of the Olympia shopping mall in Munich. 10 people, including the shooter, were killed and 36 others were injured. 

\item At 18:22 CEST, the first tweet was posted. There might be some certain delay, as we retrieve only tweets in English and the very first tweets were probably in German. The tweet is \emph{"Sadly, i think there's something terrible happening in \#Munich \#Munchen. Another Active Shooter in a mall. \#SMH"}. 

\item At 18:25 CEST, the second tweet was posted: \emph{"Terrorist attack in Munich????"}.   

\item At 18:27 CEST, traditional media (BBC) posted their first tweet. \emph{"'Shots fired' in Munich shopping centre - http://www.bbc.co.uk/news/world-europe-36870800a02026 @TraceyRemix gun crime in Germany just doubled"}.

\item At 18:31 CEST, the first misclassified tweet is posted. It was a tweet with shock sentiment and swear words: \emph{"there's now a shooter in a Munich shopping centre.. What the f*** is going on in the world. Gone mad"}. It is classified as \textit{rumor-related}.
\end{itemize}
\end{small}
We observe that at certain points in time, the volume of rumor-related tweets (for sub-events) in the event stream surges. This can lead to \textit{false positives} for techniques that model events as the aggregation of all tweet contents; that is undesired at critical moments. We trade-off this by debunking at single tweet level and let each tweet vote for the credibility of its event. We show the \emph{CreditScore} measured over time in Figure \ref{fig:munichattackCS}. It can be seen that although the credibility of some tweets are low (rumor-related), averaging still makes the \emph{CreditScore} of \textsf{Munich shooting} higher than the average of news events (hence, close to a \textit{news}). In addition, we show the feature analysis for ContainNews (percentage of URLs containing news websites) for the event \textit{Munich shooting} in Figure \ref{fig:munichattackNews}. We can see the  curve of \textit{Munich shooting} event is also close to the curve of average news, indicating the event is more news-related.

\section{Conclusion}
\label{sec:conclusion}


In this work, we propose an effective cascaded rumor detection approach using deep neural networks at tweet level in the first stage and wisdom of the  ``machines'', together with a variety of other features in the second stage, in order to enhance rumor detection performance in the early phase of an event. The proposed approach outperforms state of the 
art methods for early rumor detection. 
There is, however, still considerable room to improve the effectiveness of the rumor detection method. The 
support for events with rumor sub-events is still limited. The current model only aims not to misclassify long-running, multi-aspect events where rumors and news are mixed and evolve over time as false positive. 

\noindent\textbf{Acknowledgements.} {This work was partially funded by the German Federal Ministry of Education and Research (BMBF) under project GlycoRec (16SV7172) and K3 (13N13548).}

\newpage
\begin{subappendices}
\renewcommand{\thesection}{\Alph{section}}
\newpage
\appendix
\section{Time Period of an Event}
\label{sec:Time_Period_of_an_Event}

\begin{figure}[!h]
\centering
\subfigure[Before]{\label{fig:KKK_full}
\centering
\includegraphics[width=0.4\columnwidth]{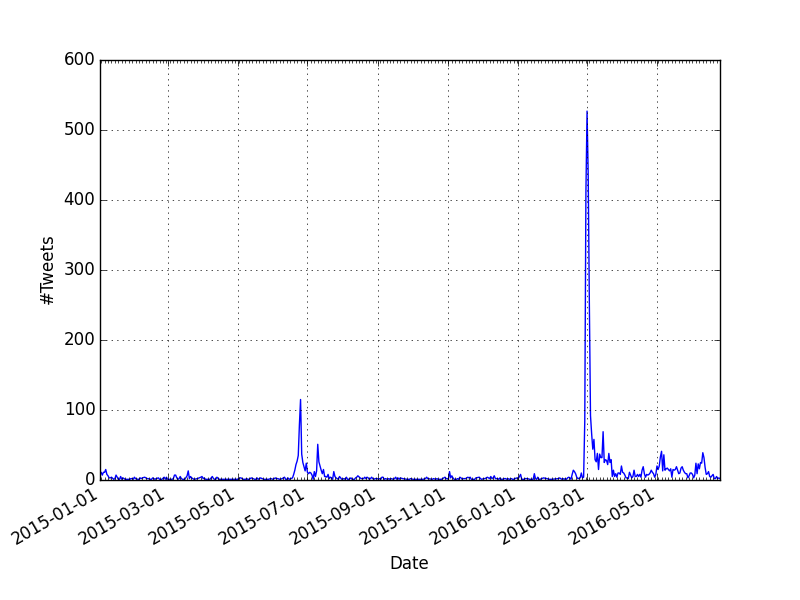}
}
\subfigure[After]{\label{fig:KKK_part}
\centering
\includegraphics[width=0.4\columnwidth]{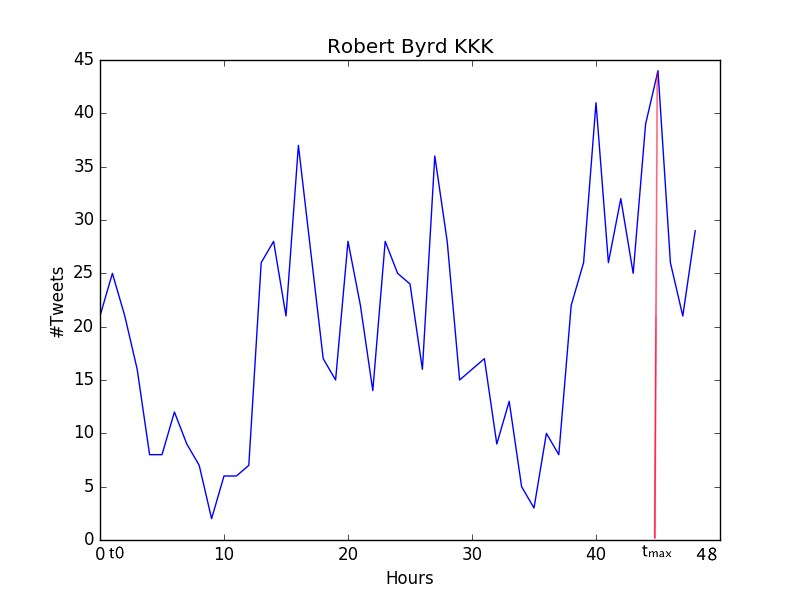}
}
\caption{tweet volume of the rumor event of Robert Byrd at full scale and after selected time period}
\label{fig:KKK_part}
\end{figure}

The time period of a rumor event is hard to define. One reason is a rumor may be created for a long time and kept existing on Twitter, but it did not attract the crowd's attention. However it can be triggered by other events after a uncertain time and suddenly spreads as a bursty event. E.g., a rumor\footnote{http://www.snopes.com/robert-byrd-kkk-photo/} claimed that Robert Byrd was member of KKK. This rumor has been circulating in Twitter for a while. As shown in Figure \ref{fig:KKK_full} that almost every day there were several tweets talking about this rumor. But this rumor was triggered by a picture about Robert Byrd kissing Hillary Clinton in 2016~\footnote{http://www.snopes.com/clinton-byrd-photo-klan/} and Twitter users suddenly noticed this rumor and it was bursted. And what we are really interested in is the tweets which are posted in hours around the bursty peak. We defined the hour with the most tweets' volume as $t_{max}$ and we want to detect the rumor event as soon as possible before its burst, so we define the time of the first tweet before $t_{max}$ within 48 hours as the beginning of this rumor event, marked as $t_{0}$. And the end time of the event is defined as $t_{end}=t_0+48$. We show the tweet volumes in Figure \ref{fig:KKK_part} of the above rumor example.
 

\section{Full Features}

\begin{table}[H]
\small
\centering
\scalebox{0.8}{
 \begin{tabular}{@{}lllllll@{}}
 \toprule
 \textbf{Category} & \textbf{Feature} & \textbf{Description}\\ \midrule
 Twitter & Hashtag & \% tweets contain \#hashtag  \cite{castillo2011information}\cite{liu2015real}\cite{qazvinian2011rumor}\cite{gupta2014tweetcred}\cite{liu2015real}\\
 	Features	& Mention &  \% tweets mention others @user  \cite{castillo2011information}\cite{liu2015real}\cite{qazvinian2011rumor}\cite{gupta2014tweetcred}\cite{liu2015real}\\
 		& NumUrls &  \# URLs in the tweet  \cite{castillo2011information}\cite{qazvinian2011rumor}\cite{gupta2014tweetcred}\cite{yang2012automatic}\cite{liu2015real}\\
 		& Retweets & average \# retweets \cite{liu2015real} \\ 
 		& IsRetweet & \% tweets are retweeted from others \cite{castillo2011information}\cite{gupta2014tweetcred}\\
 		& ContainNEWS & \% tweets contain URL and its domain's catalogue is News \cite{liu2015real}\\
 		& WotScore & average WOT score of domain in URL \cite{gupta2014tweetcred}\\
 		& URLRank5000 & \%  tweets contain URL whose domain's rank less than 5000 \cite{castillo2011information}\\
 		& ContainNewsURL & \% tweets contain URL whose domain is News Website\\
\midrule
 Text  & LengthofTweet & average tweet lengths \cite{castillo2011information}\cite{gupta2014tweetcred}\\
 Features   & NumOfChar & average \# tweet characters \cite{castillo2011information}\cite{gupta2014tweetcred}\\
   & Capital &  average fraction of characters in Uppercase \cite{castillo2011information} \\
   & Smile & \% tweets contain $:->, :-), ;->, ;-)$ \cite{castillo2011information}\cite{gupta2014tweetcred}\\
   & Sad & \%  tweets contain $:-<, :-(, ;->, ;-($ \cite{castillo2011information}\cite{gupta2014tweetcred}\\
   & NumPositiveWords & average \# positive words \cite{castillo2011information}\cite{gupta2014tweetcred}\cite{yang2012automatic}\cite{liu2015real}\\
   & NumNegativeWords & average \# negative words \cite{castillo2011information}\cite{gupta2014tweetcred}\cite{yang2012automatic}\cite{liu2015real}\\
   & PolarityScores & average polarity scores of the Tweets \cite{castillo2011information}\cite{yang2012automatic}\cite{liu2015real}\\
   & Via & \% of tweets contain via \cite{gupta2014tweetcred}\\
   & Stock & \% of tweets contain \$  \cite{castillo2011information}\cite{gupta2014tweetcred}\\
   & Question & \% of tweets contain ? \cite{castillo2011information}\cite{liu2015real}\\
   & Exclamation & \% of tweets contain ! \cite{castillo2011information}\cite{liu2015real}\\
   & QuestionExclamation & \% of tweets contain multi Question or Exclamation mark \cite{castillo2011information}\cite{liu2015real}\\ 
   & I & \% of tweets contain first pronoun like I, my, mine, we, our    \cite{castillo2011information}\cite{gupta2014tweetcred}\cite{liu2015real}\\
   & You & \% of tweets contain second pronoun like U, you, your, yours  \cite{castillo2011information}\\ 
   & HeShe & \% of tweets contain third pronoun like he, she, they, his, etc.  \cite{castillo2011information}\\ \midrule
   User & UserNumFollowers  & average number of followers \cite{castillo2011information}\cite{gupta2014tweetcred}\cite{liu2015real}\\
  Features	& UserNumFriends  & average number of friends \cite{castillo2011information}\cite{gupta2014tweetcred}\cite{liu2015real}\\
 	& UserNumTweets  & average number of users posted tweets \cite{castillo2011information}\cite{gupta2014tweetcred}\cite{yang2012automatic}\cite{liu2015real}
\\
 	& UserNumPhotos  & average number of users posted photos \cite{yang2012automatic}\\
 	& UserIsInLargeCity  & \% of users living in large city \cite{yang2012automatic}\cite{liu2015real}\\
 	& UserJoinDate & average days since users joining Twitter \cite{castillo2011information}\cite{yang2012automatic}\cite{liu2015real}
\\
 	& UserDescription  & \% of user having description \cite{castillo2011information}\cite{yang2012automatic}\cite{liu2015real}
\\
 	& UserVerified  & \% of user being a verified user\cite{yang2012automatic}\cite{liu2015real}
\\
 	& UserReputationScore & average ratio of \#Friends over (\#Followers + \#Friends) \cite{liu2015real}\\   \midrule
 Epidemiological & $\beta_{SIS}$ & Parameter $\beta$ of Model SIS \cite{jin2013epidemiological}\\
 				 Features			& $\alpha_{SIS} $ & Parameter $\alpha$ of Model SIS \cite{jin2013epidemiological}\\
 							& $\beta_{SEIZ}$ & Parameter $\beta$ of Model SEIZ \cite{jin2013epidemiological}\\
 							& $b_{SEIZ}$ & Parameter b of Model SEIZ\cite{jin2013epidemiological}\\
 							& $l_{SEIZ}$ & Parameter l of Model SEIZ \cite{jin2013epidemiological}\\
 							& $p_{SEIZ}$ & Parameter p of Model SEIZ \cite{jin2013epidemiological}\\
 							& $\varepsilon_{SEIZ}$ & Parameter $\varepsilon$ of Model SEIZ \cite{jin2013epidemiological}\\
 							& $\rho_{SEIZ}$ & Parameter $\rho$ of Model SEIZ \cite{jin2013epidemiological}\\
 							& $R_{SI}$ & Parameter $R_{SI}$ of Model SEIZ \cite{jin2013epidemiological}\\
		\midrule	
 SpikeM & $P_s$ & Parameter $P_s$ of Model Spike \cite{kwon2013prominent}\\
 			 Model 				& $P_a$ & Parameter $P_a$ of Model SpikeM \cite{kwon2013prominent}\\
 					Features		& $P_p$ & Parameter $P_p$ of Model SpikeM \cite{kwon2013prominent}\\
 							& $Q_s$  & Parameter $Q_s$ of Model SpikeM \cite{kwon2013prominent}\\
 							& $Q_a$ & Parameter $Q_a$ of Model SpikeM \cite{kwon2013prominent}\\
 							& $Q_p$ & Parameter $Q_p$ of Model SpikeM \cite{kwon2013prominent}\\ \midrule	
 Crowd Wisdom  & CrowdWisdom & \% of tweets containing "Debunking Words" \cite{liu2015real} \cite{zhao2015enquiring}\\ \midrule
 CreditScore  & CreditScore & average CreditScore\\
 \bottomrule
 \end{tabular}}
 \caption{Features of Time Series Rumor Detection Model}
 \label{tab:full_features}
\end{table}

\end{subappendices}
\bibliographystyle{abbrv}
\vspace{-0.2cm}
\begin{small}
\bibliography{websci17}
\end{small}
\end{document}